**Title: An Algorithmic Pipeline for GDPR-Compliant Healthcare Data Anonymisation: Moving Toward Standardisation**

**Authors:** Hamza Khan [1,2,3]* and Lore Menten [4]*, Liesbet M. Peeters [1, 2]

**Affiliations:**

[1] University MS Center, Biomedical Research Institute, Hasselt University, Belgium

[2] Data Science Institute (DSI), Hasselt University, Belgium

[3] The D-Lab, Department of Precision Medicine, GROW-Research Institute for Oncology and Reproduction, Maastricht University, Netherlands

[4] Faculty of Medicine and Life Sciences, Hasselt University, Belgium

* *Both authors have contributed equally to this work.*

Corresponding author: liesbet.peeters@uhasselt.be

**Background:** High-quality, real-world data (RWD) is crucial for various healthcare applications, but this data requires transformations to be shared in compliance with the General Data Protection Regulation. This regulation, with its abstract definitions regarding quasi-identifiers (QIDs) and sensitive attributes (SAs), can pose challenges to implementation.

**Objective:** This paper aims to standardise the anonymisation process of RWD to achieve GDPR compliance while preserving data utility. We propose an algorithmic approach for identifying QIDs and SAs, and we evaluate data utility in anonymised datasets by exploring and applying well-defined metrics.

**Methods:** A systematic literature review was conducted using ProQuest and PubMed to pave the way for the development of a three-stage data anonymisation pipeline consisting of "identification", "de-identification", and "quasi-identifier dimension evaluation". The pipeline was further implemented, validated and tested on mock datasets designated to resemble RWD. Privacy was assessed using metrics including k-anonymity, ℓ-diversity, and t-closeness, while data utility was evaluated using non-uniform entropy (NUE).

**Results:** The SLR identified two relevant publications for QID and SA identification and five for data utility metrics. The pipeline's stages were successfully applied to two mock datasets, containing 500 and 1000 rows, respectively. In the identification stage, attributes were classified based on re-identification risk with α and β thresholds, set at 25% and 1% for the 500-row dataset, and 10% and 1% for the 1000-row dataset. Privacy metrics showed k-anonymity improved from one to four and one to 110 for the 500-row and 1000-row datasets, respectively. NUE scores were similar across datasets, with 69.26% and 69.05%, respectively, indicating consistent data utility.

**Conclusion:** This paper provides a GDPR-compliant anonymisation pipeline for healthcare data, offering a systematic, reproducible approach to attribute identification and utility evaluation. The publicly available code supports the standardisation of the anonymisation process, contributing to both data privacy and open science.

# Introduction

Real-world data (RWD) has recently gained attention and is defined in multiple ways depending on the context. In this paper, RWD refers to health data routinely gathered from different sources within healthcare services, as opposed to data collected in experimental settings. Disease registries contribute to the accumulation of RWD by directly collecting information from patients or aggregating data extracted from electronic health record systems. In some cases, a blend of both methods may be used to ensure comprehensive data collection [1].

High-quality RWD is essential in patient care, quality improvement, safety monitoring and research [1]. Sharing data enhances confidence and trust in research findings, enables reproducibility and promotes the exploration of new hypotheses. It maximises progress by preventing duplication and utilising insights from each trial. Additionally, it satisfies ethical responsibilities towards participants and benefits many stakeholders. As awareness regarding its importance grows, numerous global initiatives now advocate for medical data sharing. These efforts pave the way for open science while safeguarding patient privacy [2]. Statistical Disclosure Control (SDC) is a field that quantifies and limits the risk of re-identifying individuals in released micro-data and therefore recommends that data sharing be guided by formal, quantitative risk-and-utility metrics [3]. However, sharing RWD also presents significant challenges, particularly in ensuring compliance with privacy regulations such as the General Data Protection Regulation (GDPR) [4].

When discussing data sharing, the GDPR inevitably enters the conversation. The GDPR stands as a cornerstone in safeguarding data privacy, forming a framework for managing, processing, and protecting personal data [4]. To share data in a GDPR-compliant manner, a dataset must undergo transformations, as shown in Figure 1. Typically, this process involves identifying attributes into three categories:

- **Direct identifiers (DIDs):** Attributes that directly identify an individual, such as a full name or a social security number [5].

- **Quasi-identifiers (QIDs):** Also known as indirect identifiers, are attributes that do not uniquely identify an individual on their own but, when combined with other information, could be used to re-identify individuals, such as date of birth, gender, or postal code [5].

- **Sensitive attributes (SAs):** Attributes that contain sensitive information about an individual, such as medical conditions, income, or ethnicity [5].

After categorising the attributes, de-identification strategies such as masking, generalisation and suppression are applied, followed by anonymisation techniques [6]. For the sake of this paper, we reserve the term "de-identification" for the transformation step and "anonymisation" for the resulting dataset that meets the agreed disclosure-risk thresholds, in line with the modern SDC terminology. Once anonymised, a bias assessment check can be performed to assess data usefulness.

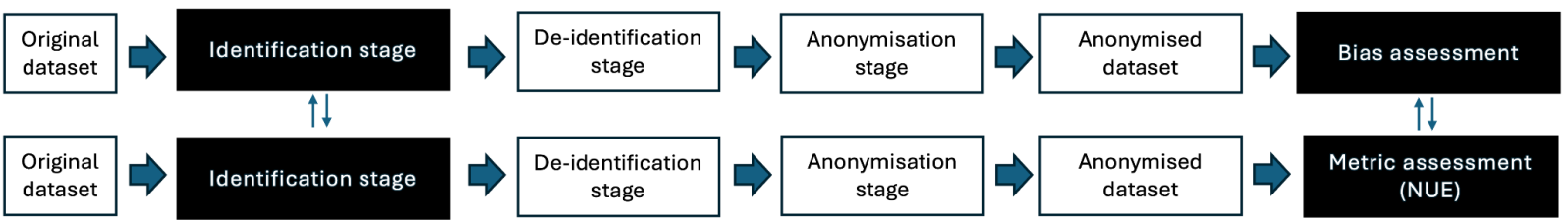

Figure 1: The data anonymisation process and contribution of this paper. *NUE = Non-Uniform Entropy*

Although GDPR is critical for ensuring patient privacy, it also presents significant challenges for researchers and healthcare professionals aiming to utilise RWD effectively [6]. This challenge is particularly evident during the initial identification stage, where definitions for quasi-identifiers (QIDs) and sensitive attributes (SAs) remain vague. A research gap exists within the identification stage due to the abstract nature of definitions for QIDs and SAs. Specifically within the context of healthcare data, where a dataset can have multiple QIDs that can lead to the re-identification of individuals when cross-referenced with other data sources [7,8]. The ambiguity in GDPR's definitions of QIDs, DIDs, and SAs can lead to multiple interpretations, making implementation inconsistent and challenging for organisations to achieve compliance [9–12].

To address these concerns, this paper aims to initiate the standardisation of the anonymisation process in order to streamline data sharing efforts. First, an algorithmic approach based on mathematical logic will be used for the identification of attributes, reducing the reliance on subjective GDPR interpretations. Second, additional data usefulness metrics will be identified. Figure 1 presents the contributions of this paper.

These aims, therefore, translate into the following research questions:

- RQ1: Given the already present definitions entailing quasi-identifiers and sensitive attributes, how can the identification process for these types of attributes be standardised using an algorithmic approach?

- RQ2: Which metric should be used to evaluate data usefulness of an anonymised dataset?

By addressing these questions, this paper seeks to advance both the methodology and practical application of RWD anonymisation, ultimately promoting open science without compromising patient privacy.

## Methods

This paper bridges healthcare and computer science research, requiring two methodologies to balance a strong design with practical significance. It starts with a systematic literature review (SLR), conducted and reported according to the Preferred Reporting Items for Systematic Reviews and Meta-Analyses (PRISMA) guideline [13], using the Parsif.al tool to facilitate the review process [14,15]. The findings from the SLR were then integrated into the development of a new data anonymisation pipeline. Subsequently, the pipeline was then transformed into code, validated and tested on two datasets. Finally, the results were reported using selected sections from the Standards for Reporting of Diagnostic Accuracy (STARD) guidelines [16].

### Systematic Literature Review (SLR)

*Inclusion and exclusion criteria*

To maintain focus on the relevant literature, publications were required to address structured and tabular data due to the scope of this paper. Only peer-reviewed articles published in English were included to ensure quality and accessibility, while books were excluded to mitigate potential accessibility issues.

*Search strategy*

Publications were gathered from ProQuest and PubMed between December 25$^{th}$ of 2023, and February 11$^{th}$ of 2024. These databases were chosen due to their extensive biomedical and informatics literature collection, ensuring comprehensive coverage of relevant topics. Because there are two research questions, i.e., RQ1 and RQ2, two search strings were composed (Table 1). Field codes were applied to limit the search to publications where search terms appeared in the title or the abstract. No Medical Subject Headings (MeSH-terms) were used to reduce the possibility of missing recent publications. Filters limited the results to peer-reviewed publications written in English while excluding books.

Table 1: Search strings

|  | **RQ1: Given the already present definitions entailing quasi-identifiers and sensitive attributes, how can the identification process for these types of attributes be standardised using an algorithmic approach?** | **RQ2: What are the methods used to evaluate data usefulness of an anonymised dataset?** |
|---|---|---|
| **ProQuest** | TIAB(("quasi-identifier" OR "QID" OR "QI") AND ("recogni*" OR "classif*" OR "detect*" OR "discover*" OR "identif*" OR "find*" OR "solv*") AND ("algorithm") AND ("privacy")) | TIAB(("anonymi*") AND ("metric" OR "assess*" OR "evaluat*" OR "measur*") AND ("data usefulness" OR "data quality" OR "data utility")) |
| **PubMed** | ((quasi-identifier[Title/Abstract]) OR (QID[Title/Abstract]) OR (QI[Title/Abstract])) AND ((recogni*[Title/Abstract]) OR (classif*[Title/Abstract]) OR (detect*[Title/Abstract]) OR (discover*[Title/Abstract]) OR (identif*[Title/Abstract]) OR (find*[Title/Abstract]) OR (solv*[Title/Abstract])) AND (algorithm[Title/Abstract]) AND (privacy[Title/Abstract]) | (anonymi*[Title/Abstract]) AND ((metric[Title/Abstract]) OR (assess*[Title/Abstract]) OR (evaluat*[Title/Abstract]) OR (measur*[Title/Abstract])) AND ((data usefulness[Title/Abstract]) OR (data quality[Title/Abstract]) OR (data utility[Title/Abstract])) |

*Selection and management of the publications*

The selection process was documented in a flowchart (Figure 2). Publications were initially screened on title and abstract, focusing on identifying publications relevant to the research objectives. If the publications aligned with the research objective, full texts were reviewed. EndNote was used to effectively manage the publications [17].

*Data items and data collection*

The first outcome included an algorithmic approach to identify attributes, while any metrics for data usefulness served as a second outcome. Appendix 1 contains the data extraction forms created to facilitate the data extraction process.

*Quality assessment*

The quality of the included publications was assessed using quality assessment checklists (Appendix 2). Since the two research questions targeted different kinds of publications, two separate checklists were used. In the case of an absence of validated checklists for the type of publications in question, custom checklists were developed based on available literature [18].

## Dataset

*Experimental dataset*

The experimental dataset was a mock dataset designed to closely resemble RWD. It consisted of 17 attributes or columns and was created with two versions, containing 500 and 1000 rows each [6]. The datasets were generated using a mock data generator to analyse the effect of dataset size on the privacy and utility metrics (described below). The full description of the datasets is available in the [GitHub repository](#).

*Handling Missing data*

Since RWD is often characterised by low data quality, a function was designed to handle missing values [19]. Missing values were replaced with the string "missing", allowing for the calculation of the percentage of missing values in an attribute. If more than 85% of an attribute's data was missing, that attribute was dropped to minimise noise. This strategy allowed for extended matching within k-anonymity, where missing values could be matched with other missing values, unlike a basic match that does not allow missing values to match even with other missing values [20].

## Pipeline

Following the insights gathered from the SLR, a pipeline was developed to address the challenges in data anonymisation stemming from the ambiguity pertaining to the GDPR definitions of QIDs and SAs. The pipeline, as depicted in Figure 2, was built around three major stages: the identification stage, the de-identification stage, and the quasi-identifier dimension stage. The code for the pipeline can be consulted on the [GitHub repository](#) mentioned in the code availability section.

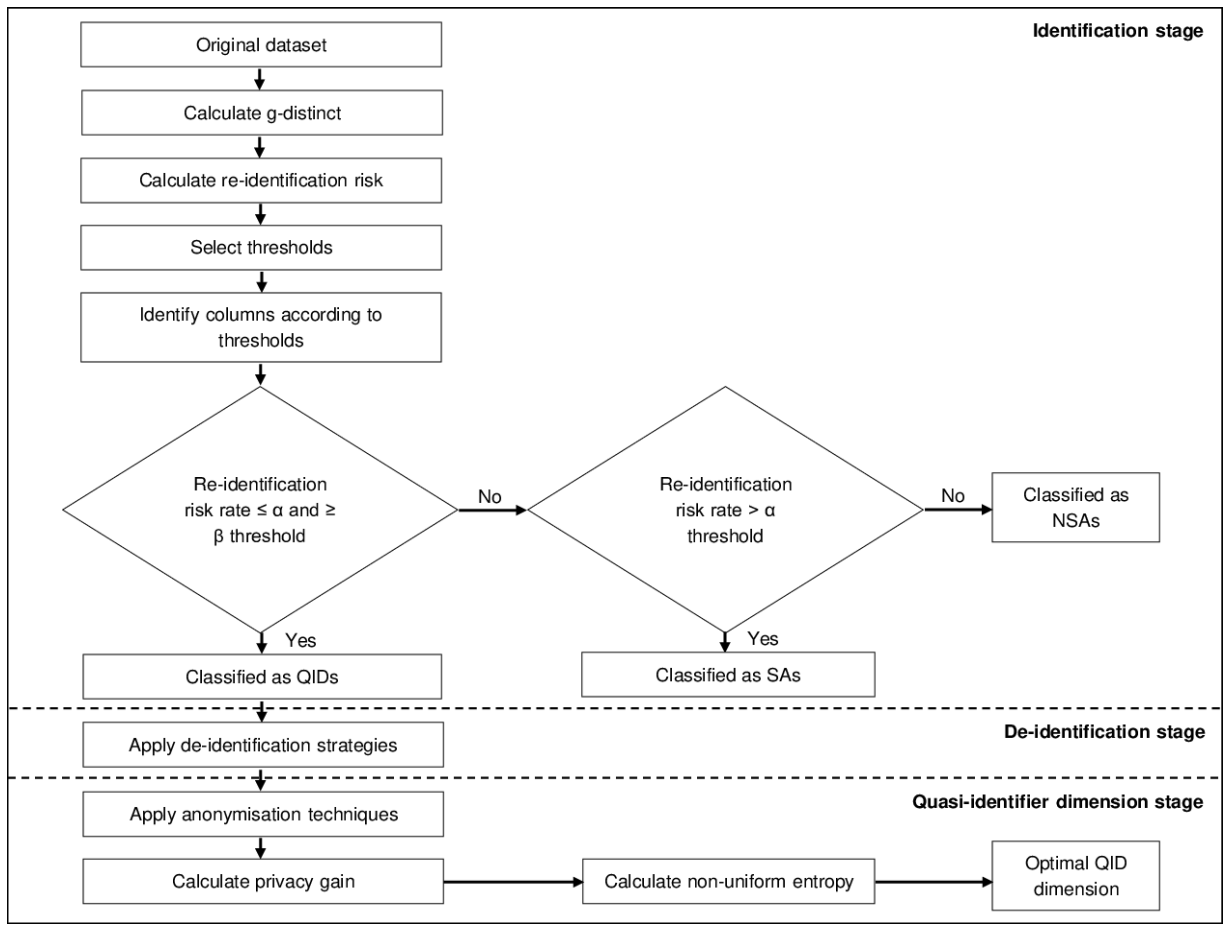

Figure 2: Flowchart of the pipeline

*Identification stage*

The identification stage focused on assessing re-identification risks using an algorithmic approach. Three steps for the identification stage were identified: calculate g-distinct, calculate re-identification risk, and classify attributes according to re-identification risk thresholds [21,22]. The calculation of g-distinct values is based on attribute uniqueness: each g-distinct value represents how unique a value is within an attribute. Based on these g-distinct values, the re-identification risk rate of every attribute can be calculated. This rate is determined by the sum of all g-distinct values within an attribute. Subsequently, α and β thresholds are established to classify attributes as QIDs, SAs or non-sensitive attributes (NSAs). Attributes surpassing the α threshold are labelled as SAs. Attributes with a rate lower than or equal to α but higher than or equal to β are considered QIDs. The remaining attributes with a rate below β are classified as NSAs [21]. The classified attributes serve as the output of this stage.

*De-identification stage*

The de-identification stage implemented commonly used privacy protection strategies such as suppression, masking, generalisation and aggregation. Suppression involves removing values or attributes, which maximises privacy protection but often reduces data usefulness. Health data protection standards often necessitate some degree of suppression [23,24]. Masking obscures data by replacing a value or a part of it with placeholders, making it difficult to retrieve the original information. Generalisation reduces information specificity by representing values within broader ranges, thereby decreasing granularity but also reducing data utility due to information loss [23,24]. Aggregation involves grouping together raw data, which allows the release of summary statistics or information about small groups within datasets, rather than revealing entire datasets [23]. These strategies were applied iteratively, starting with the QIDs posing the highest risk. This systematic approach ensures that attributes with a higher likelihood of re-identification were anonymised first, thereby maximising the protection of sensitive data.

*Quasi-identifier dimension stage*

The quasi-identifier dimension stage evaluated the privacy and usefulness of the de-identified dataset by applying privacy metrics and usefulness metrics (see below) to ensure an optimal balance. The evaluation process aimed to maintain privacy without significantly compromising data utility.

Privacy metrics

The privacy metrics used to assess the level of privacy achieved in the de-identified dataset are as follows:

- k-anonymity: Ensures that each record in the dataset is indistinguishable from at least k-1 other records based on a combination of QIDs, preventing identity disclosure. The group of indistinguishable records is called an equivalence class [23–26].

- ℓ-diversity: Addresses vulnerabilities in k-anonymous datasets, such as homogeneity and background knowledge attacks. It ensures that each equivalence class contains at least ℓ well-represented values for the sensitive attribute to protect against inferential disclosure [24,27].

- t-closeness: Mitigates risks associated with ℓ-diversity, such as skewness and similarity attacks, by ensuring that the distribution of sensitive attributes within an equivalence class is within a threshold (t) of the distribution across the entire dataset [28].

Usefulness metrics

To evaluate the usefulness of the de-identified dataset, several usefulness metrics were identified from the five included publications addressing the second research question [29–33]. Metrics considered included non-uniform entropy (NUE), utility criterion, and clustering. NUE received the best results for general-purpose usage and was the most comprehensive and well-documented approach among the three [29,32]. Hence, NUE was used in this paper as the only metric to assess data utility. NUE quantifies the information loss in the de-identified dataset by comparing the frequency distributions of attribute values in the original and anonymised datasets [29,32]. An inverse NUE metric was also calculated to reflect retained data utility.

Overall, we define the optimal QID dimension as the smallest subset of QIDs that (i) achieves $k \geq 2$, $\ell \geq 2$ and $t \leq 0.8$, and (ii) maximises NUE. Based on the evaluation of these privacy and usefulness metrics, an optimal selection of QIDs was determined [21]. The selection process, among others, ensured that the k-anonymity level of at least two was maintained, thereby avoiding situations in which individuals could be easily identified.

*Validation*

The identification stage in the state-of-the-art example was outsourced, and the original dataset was no longer accessible, making it difficult to validate the pipeline [6]. However, where the identified publications provided sufficient transparency regarding their experiments, the pipeline was validated with datasets from these papers [21].

# Results

This section presents the findings in a structured manner, beginning with insights from the systematic literature review, followed by the implementation and evaluation of the anonymisation pipeline. The selection process has been based on the PRISMA flowchart [13].

## Systematic Literature Review Findings

As shown in Figure 3, the SLR aimed at addressing RQ1 and RQ2. Regarding RQ1, a total of twenty relevant publications were identified. After going through the screening process, only two articles were selected that provided insights into algorithmic identification methods for QIDs and SAs (Figure 3). Furthermore, with regards to scoring, both articles scored five out of eight due to not incorporating machine learning and lacking a code repository.

For RQ2, a total of 162 articles were identified, and post selection process, the number came down to five. The five final articles received divergent scores. The lowest score was zero out of five. Three articles scored three out of five due to not validating the methodology and the absence of a code repository. One article provided a repository, resulting in the highest score of four out of five. More information on the data extraction forms for the included publications can be found in Appendix 5, while Appendix 6 shows the quality assessment checklists for these publications.

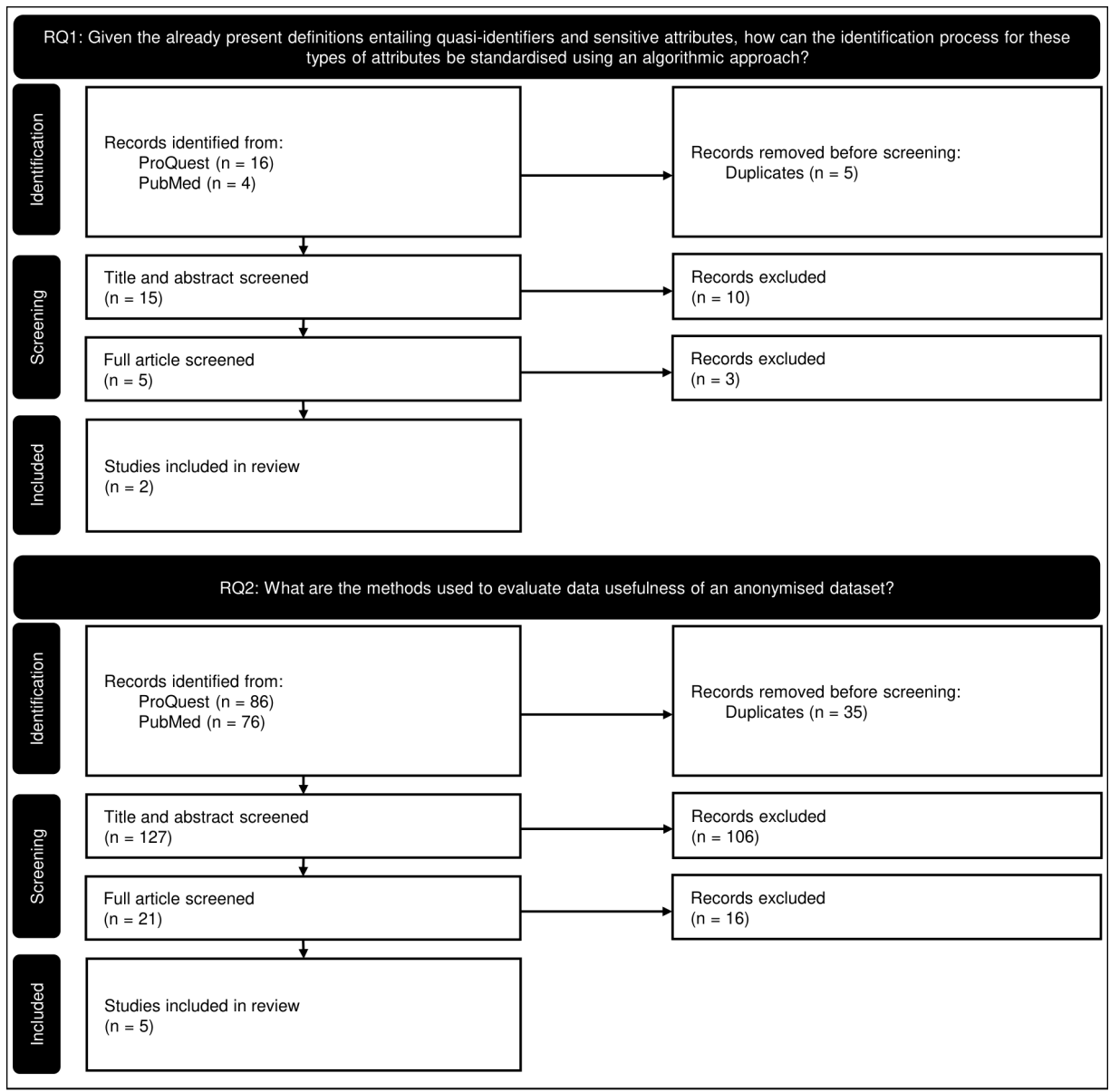

Figure 3: Flowchart of the selection process

Pipeline implementation and evaluation

With regards to the pipeline, three main steps of the pipeline consisting of the "identification stage", the "de-identification stage", and the "QID dimension stage" were executed (Figure 2) [21]. The code for the pipeline has been made available in a [GitHub repository](GitHub repository).

*Identification Stage*

The output of the identification stage consisted of re-identification risks for each attribute and the classification of these attributes according to the re-identification risk thresholds (Appendix 7). For the first dataset, which contained 500 rows, the α threshold was set to 25% and the β threshold to 1%. As for the second dataset with 1000 rows, α was set to 10% and β to 1%. The attribute "covid19_self_isolation" was excluded because the missing values surpassed the predefined threshold of 85%, reaching 88% in the first dataset and 91.8% in the second dataset. The attribute "secret_name" was also suppressed since this was a DID.

*The de-identification and QID dimension stage*

The QID dimension stage was executed after the de-identification of QIDs and SAs in both datasets. For the first dataset, the optimal QID dimension was five, indicating that all QIDs needed to be de-identified. This was primarily because k-anonymity for the other dimensions was below two. For this dimension, k-anonymity was four, ℓ-diversity was two for both SAs and t-closeness was 0.74. K-anonymity before anonymisation was one, resulting in a privacy gain of three. NUE was 69.26%, and the inverse NUE was 30.74%.

The optimal QID dimension in the second dataset was three, implying that all three QIDs should undergo de-identification. K-anonymity was six with two de-identified QIDs, satisfying the minimum requirement of two to ensure that no person remains unique. This resulted in a privacy gain of five. ℓ-diversity varied across QIDs, with "bmi" and "ms_diagnosis_date" having a value of three and "edss" having a value of two. The t-closeness value was 0.61. NUE was 53.61,% and the inverse NUE was 46.39%. After de-identifying the third QID, k-anonymity improved from 1 to 110, resulting in a privacy gain of 109. ℓ-diversity for "bmi", "ms_diagnosis_date" and "edss" was three, six and two, respectively, while t-closeness decreased to 0.32. NUE was 69.05%, with an inverse NUE of 30.95%.

**Discussion**

This study provides a systematic approach to developing a GDPR-compliant data anonymisation pipeline tailored for healthcare data sharing. The findings address significant gaps in the current literature, particularly the lack of standardised methods for identifying QIDs and SAs. The SLR showed that most existing approaches lack machine learning integration and public code repositories, creating a gap in reproducibility and robustness. By developing an open-source tool and making it publicly available, this paper contributes towards an open science culture, which is essential for promoting transparency and collaboration.

The study identified a method for attribute identification based on re-identification risk, with NUE as a well-supported metric for data usefulness. Re-identification risk seems just one of countless methods to identify attributes, with every method claiming superiority over the others by minimising information loss or supposedly being more efficient [34–37]. However, a common drawback among these methods is the lack of publicly available code, exemplifying that these publications fail to contribute to an open science culture. While most methods provide pseudocode, this limits usability as they still have to be translated into the desired programming language. During the validation of the code, the results were not fully identical to those in the source paper. However, this source paper had methodological flaws. Firstly, the β threshold was set to zero, requiring NSAs to be negative. This seemed highly unlikely as risk rates are strictly positive. Additionally, some steps had limited explanation, such as converting the re-identification risk rate to a percentage [21].

The absence of a code repository in the source paper necessitated the development of code from scratch, except for the de-identification and anonymisation strategies outlined in the methods. While the results of the experiment were reported in sufficient detail to use as validation for the code [21], relying solely on pseudocode and definitions can lead to inconsistencies. When tested with a dataset from the source paper, the re-identification risk rates were similar, though classification using the same thresholds yielded slightly different results.

For usefulness metrics, numerous alternatives were also identified during the SLR, with one ultimately implemented. Unlike attribute identification methods, usefulness metrics are well-documented and publicly available. An example of existing open-source software is ARX [38]. It supports a variety of privacy models, data transformation models, and utility and risk analysis techniques. They provide code to compute re-identification risks for various attacker models and numerous data quality models. However, all these methods are stored in individual files in their repository, resulting in a lack of cohesion compared to the pipeline developed in this paper. Another major weakness is that ARX is only available in Java. Python is the programming language of choice for data scientists and developers in data analysis and numerical computations [39,40]. Consequently, the software is less usable in these specific fields.

The experiment shows that as the dataset size increases, the re-identification risk decreases. This is logical, as values lose their uniqueness when they are present in a larger number of records. Consequently, smaller datasets inherently carry a greater re-identification risk, necessitating stronger de-identification strategies to satisfy required anonymity levels. This leads to a decrease in overall utility when weighed against the privacy gain [41]. In both datasets, NUE was similar for the best QID dimensions, with 69.26% in the 500 row dataset and 69.05% in the 1000 row dataset. However, the privacy gain in the 500 row dataset was merely 3, whereas the 1000 row dataset achieved a privacy gain of 109. Therefore, even though utility remains comparable between the two datasets, the privacy gain differs greatly.

A major strength of this paper is its methodology. The incorporation of SLR components contributes to the quality of the methodology, for instance, the use of a predefined search string across preselected databases and quality assessments of included publications. This systematic search for literature also provides a solid scientific foundation for the pipeline. The experiment further enhances the significance of the findings within the practical domain. Additionally, the reporting quality is elevated by the transparent documentation of the decisions made. In the SLR itself, the inclusion of only peer-reviewed publications enhances the reliability of findings, with most selected publications scoring well on quality assessments, primarily losing points due to missing code repositories or lack of methodological validation.

The greatest asset of this paper is its contribution to the open science culture. The code is available on [GitHub](), complete with comprehensive documentation and mock datasets. The inclusion of mock datasets that mimic real privacy risks without compromising actual privacy provides a safe way to develop skills in handling RWD. This significantly enhances the educational value of this paper. The use of realistic datasets also ensures that the algorithm can effectively handle low-quality data often found in such datasets [19]. Moreover, the code can be applied to any dataset, making it an invaluable tool for people working with data anonymisation. The QID dimension stage is designed to accept datasets and calculate necessary parameters for comparing them. With some adjustments, it could compare two datasets where the same QID is de-identified differently, allowing users to determine the most effective de-identification strategy. The code was also developed with user-friendliness in mind, utilising user input prompts instead of requiring manual code changes.

For the SLR, an important limitation is the limited amount of publications retrieved by the search strings. However, this was to be expected considering the relatively recent emergence of privacy regulations like the GDPR. Regardless, it is essential to reflect on whether this is a result of flaws in the search strings or in the article selection process. Another flaw is that almost all steps were executed by a single individual, which is also to be expected in the context of a paper. Ideally, an SLR is executed by at least two researchers.

As for the pipeline, the arbitrary selection of α and β thresholds poses a major challenge to the objectivity of the results. The idea that SAs have a higher re-identification risk than QIDs should be interpreted with caution, since this assumption has not been thoroughly tested. The fact that there is no consensus about where these thresholds should be proves again that the definitions of QIDs and SAs are not clear enough.

Another important limitation concerns low-cardinality variables. In both mock datasets, sex and the binary covid19_diagnosis flag fell below the β-threshold (g-distinct ≈ 0.4 %) and were therefore labelled non-sensitive. While this is correct under a purely data-driven rule, an adversary could readily know such information, and many disclosure-control frameworks would treat these attributes as quasi-identifiers by default [25,42]. We chose not to override the automated label so the experiment illustrates a fully algorithmic workflow, but we acknowledge that future versions should allow users to force selected variables into the QID set or weight g-distinct scores by attacker knowledge.

Moreover, further research is necessary to develop more objective methods for selecting re-identification risk thresholds. Regardless, the risk rates can be used to guide the order in which attributes should be de-identified. While measures of usefulness have been extensively documented, greater emphasis should be placed on creating comprehensive pipelines that integrate all steps of the anonymisation process, rather than solely focusing on generating code for individual usefulness measures.

Additionally, open-source tools should not be considered a nice-to-have but a necessity for publications in this field of study. The publication of code should be encouraged, since this promotes reproducibility, enhances usability, and contributes to the open science culture. Specific research in healthcare informatics should be supported, with a focus on combining methodologies to ensure robust designs together with the development of open-source tools.

## Conclusion

The results of this paper are promising, demonstrating that objective ways to identify attributes exist and various data usefulness metrics are available. The drawback is not the scarcity of methods but rather the absence of open-source tools to apply them. In this regard, this paper makes a significant contribution by providing a tool developed through a robust, systematic methodology. The code is adaptable to various datasets and can be easily customised with minimal adjustments, making it a versatile resource for various applications.

## Code availability

The pipeline was developed using Python and split up into two files. The first file contains the identification stage, while the second file contains the QID dimension stage. The Python libraries used are pandas 2.1.4, NumPy 1.26.2 and pyCANON 1.0.1.post2. The datasets from the experiment are provided in a CSV format. Users can access the GitHub repository [here](#).

## Acknowledgements

One of the authors, Hamza Khan, a PhD researcher at Hasselt University and Maastricht University, gratefully acknowledges funding support from the Bijzonder Onderzoeksfonds (BOF), a Special Research Fund dedicated to supporting fundamental research at Flemish universities in Belgium.

## Conflict of interest

Hamza Khan, Lore Menten, and Liesbet Peeters declare that they have no conflicts of interest related to this research.

# Appendices

## Appendix 1: Data extraction forms

| Given the already present definitions entailing quasi-identifiers and sensitive attributes, how can the identification process for these types of attributes be standardised using an algorithmic approach? | What are the methods used to evaluate data usefulness of an anonymised dataset? |
|---|---|
| Title | Title |
| Authors | Authors |
| Publication year | Publication year |
| Objectives | Objectives |
| Methodology | Methodology |
| Performance measures | Key takeaways |
| Key takeaways | |

# Appendix 2: Quality assessment checklists

| Given the already present definitions entailing quasi-identifiers and sensitive attributes, how can the identification process for these types of attributes be standardised using an algorithmic approach? | What are the methods used to evaluate data usefulness of an anonymised dataset? |
|---|---|
| Was there a clear description of the aims and purposes of the research? | Was there a clear description of the aims and purposes of the research? |
| Was the algorithm clearly described (e.g. flowchart, pseudocode, ...) | Was the experimental dataset described? |
| Was the experimental dataset described? | Were any metrics used to validate the methodology? |
| Were any metrics used to validate the methodology? | Were the metrics clearly described? |
| Was the quality of the anonymised data assessed? | Is there a repository of the code? |
| Was the quality assessment done using simple statistical methods or machine learning? | |
| Were there any hyperparameters that were finetuned? | |
| Is there a repository of the code? | |

# Appendix 3: De-identification strategies

## Suppression

Original data:

| Name | Postal code | Age | Sex | Diagnosis |
|---|---|---|---|---|
| Emma | 3500 | 25 | F | Gastric flu |
| Bob | 3510 | 32 | M | Flu |
| Tommy | 3520 | 36 | M | COVID |
| Michael | 3530 | 45 | M | Gastric flu |
| Sara | 3540 | 23 | F | Flu |
| Ziggy | 3550 | 43 | M | COVID |

After suppression:

| Name | Postal code | Age | Sex | Diagnosis |
|---|---|---|---|---|
|  | 3500 | 25 | F | Gastric flu |
|  | 3510 | 32 | M | Flu |
|  | 3520 | 36 | M | COVID |
|  | 3530 | 45 | M | Gastric flu |
|  | 3540 | 23 | F | Flu |
|  | 3550 | 43 | M | COVID |

## Masking

Original data:

| Name | Postal code | Age | Sex | Diagnosis |
|---|---|---|---|---|
| Emma | 3500 | 25 | F | Gastric flu |
| Bob | 3510 | 32 | M | Flu |
| Tommy | 3520 | 36 | M | COVID |
| Michael | 3530 | 45 | M | Gastric flu |
| Sara | 3540 | 23 | F | Flu |
| Ziggy | 3550 | 43 | M | COVID |

After masking:

| Name | Postal code | Age | Sex | Diagnosis |
|---|---|---|---|---|
| xxxx | 3500 | 25 | F | Gastric flu |
| xxxx | 3510 | 32 | M | Flu |
| xxxx | 3520 | 36 | M | COVID |
| xxxx | 3530 | 45 | M | Gastric flu |
| xxxx | 3540 | 23 | F | Flu |
| xxxx | 3550 | 43 | M | COVID |

## Generalisation

Original data:

| Name | Postal code | Age | Sex | Diagnosis |
|---|---|---|---|---|
| Emma | 3500 | 25 | F | Gastric flu |
| Bob | 3510 | 32 | M | Flu |
| Tommy | 3520 | 36 | M | COVID |
| Michael | 3530 | 45 | M | Gastric flu |
| Sara | 3540 | 23 | F | Flu |
| Ziggy | 3550 | 43 | M | COVID |

After generalisation:

| Name | Postal code | Age | Sex | Diagnosis |
|---|---|---|---|---|
| Emma | 3500 | 20-29 | F | Gastric flu |
| Bob | 3510 | 30-39 | M | Flu |
| Tommy | 3520 | 30-39 | M | COVID |
| Michael | 3530 | 40-49 | M | Gastric flu |
| Sara | 3540 | 20-29 | F | Flu |
| Ziggy | 3550 | 40-49 | M | COVID |

## Aggregation

Original data:

| Name | Postal code | Age | Sex | Diagnosis |
|---|---|---|---|---|
| Emma | 3500 | 25 | F | Gastric flu |
| Bob | 3510 | 32 | M | Flu |
| Tommy | 3520 | 36 | M | COVID |
| Michael | 3530 | 45 | M | Gastric flu |
| Sara | 3540 | 23 | F | Flu |
| Ziggy | 3550 | 43 | M | COVID |

After aggregation:

| Name | Postal code | Age | Sex | Diagnosis |
|---|---|---|---|---|
| Emma | 3500 | 25 | F | Digestive |
| Bob | 3510 | 32 | M | Respiratory |
| Tommy | 3520 | 36 | M | Respiratory |
| Michael | 3530 | 45 | M | Digestive |
| Sara | 3540 | 23 | F | Respiratory |
| Ziggy | 3550 | 43 | M | Respiratory |

## Appendix 4: Anonymisation techniques (adapted from [5])

| Direct identifier | Quasi-identifier | Quasi-identifier | Quasi-identifier | Quasi-identifier | Quasi-identifier | Sensitive attribute | Sensitive attribute |
|---|---|---|---|---|---|---|---|
| secret_name | age | edss | covid19_symptoms | comorbidities | ms_type | ms_diagnosis_date | bmi |
| C_1038 | 38 | 0,7 | shortness_breath | no | CIS | 1997 | 23,8 |
| P_311 | 26 | 3,7 | pneumonia | no | not_sure | 2017 | 28,4 |
| C_1060 | 38 | 1,3 | chills | no | not_sure | 2005 | 32,1 |
| P_289 | 25 | 2,2 | chills | chronic_kidney_disease | CIS | 2012 | 24,7 |
| P_338 | 41 | 0,5 | no | chronic_liver_disease | PPMS | 2015 | 30,7 |
| C_1174 | 64 | 3,5 | no | lung_disease | PPMS | 2020 | 20 |
| C_1349 | 53 | 5,5 | no | other | RRMS | 1990 | 21,6 |
| P_512 | 46 | 6,2 | no | cardiovascular_disease | RRMS | 2013 | 27,4 |

De-identification

| ~~Direct identifier~~ | Quasi-identifier | Quasi-identifier | Quasi-identifier | Quasi-identifier | Quasi-identifier | Sensitive attribute | Sensitive attribute |
|---|---|---|---|---|---|---|---|
| ~~secret_name~~ | age | edss | covid19_symptoms | comorbidities | ms_type | ms_diagnosis_date | bmi |
| ~~C_1038~~ | 18-40 | 0.0-4.5 | yes | no | other | 1995-1999 | healthy weight |
| ~~P_311~~ | 18-40 | 0.0-4.5 | yes | no | other | 2015-2019 | overweight |
| ~~C_1060~~ | 18-40 | 0.0-4.5 | yes | yes | other | 2005-2009 | obese |
| ~~P_289~~ | 18-40 | 0.0-4.5 | yes | yes | other | 2010-2014 | healthy weight |
| ~~P_338~~ | 40-69 | 0.0-4.5 | no | yes | progressive_MS | 2015-2019 | obese |
| ~~C_1174~~ | 40-69 | 0.0-4.5 | no | yes | progressive_MS | > 2019 | healthy weight |
| ~~C_1349~~ | 40-69 | 5.0-10.0 | no | yes | relapsing_remitting | 1990-1994 | healthy weight |
| ~~P_512~~ | 40-69 | 5.0-10.0 | no | yes | relapsing_remitting | 2010-2014 | overweight |

k-anonymity and ℓ-diversity

| ~~Direct identifier~~ | Quasi-identifier | Quasi-identifier | Quasi-identifier | Quasi-identifier | Quasi-identifier | Sensitive attribute | Sensitive attribute |
|---|---|---|---|---|---|---|---|
| ~~secret_name~~ | age | edss | covid19_symptoms | comorbidities | ms_type | ms_diagnosis_date | bmi |
| ~~C_1038~~ | 18-40 | 0.0-4.5 | yes | no | other | 1995-1999 | healthy weight |
| ~~P_311~~ | 18-40 | 0.0-4.5 | yes | no | other | 2015-2019 | overweight |
| ~~C_1060~~ | 18-40 | 0.0-4.5 | yes | yes | other | 2005-2009 | obese |
| ~~P_289~~ | 18-40 | 0.0-4.5 | yes | yes | other | 2010-2014 | healthy weight |
| ~~P_338~~ | 40-69 | 0.0-4.5 | no | yes | progressive_MS | 2015-2019 | obese |
| ~~C_1174~~ | 40-69 | 0.0-4.5 | no | yes | progressive_MS | > 2019 | healthy weight |
| ~~C_1349~~ | 40-69 | 5.0-10.0 | no | yes | relapsing_remitting | 1990-1994 | healthy weight |
| ~~P_512~~ | 40-69 | 5.0-10.0 | no | yes | relapsing_remitting | 2010-2014 | overweight |

## Appendix 5: Filled out data extraction forms

| | Given the already present definitions entailing quasi-identifiers and sensitive attributes, how can the identification process for these types of attributes be standardised using an algorithmic approach? | |
|---|---|---|
| Title | Quasi-identifier recognition algorithm for privacy preservation of cloud data based on risk reidentification [22] | Quasi-identifier recognition with echo chamber optimization-based anonymization for privacy preservation of cloud storage [21] |
| Authors | Mansour HO, Siraj MM, Ghaleb FA, Saeed F, Alkhammash EH, Maarof MA | Jadhav PS, Borkar GM |
| Publication year | 2021 | 2024 |
| Objectives | - to overcome the identity disclosure resulting from QID linking<br>- to reduce the leakage of privacy by proposing a QID recognition algorithm based on risk rate reidentification | - to identify the quasi-attributes based on clustering<br>- to maintain privacy preservation in the cloud based on the echo chamber optimization as well as the optimized k-anonymisation process |
| Methodology | - data preprocessing<br>- compute risk rate for all attributes<br>- select classification thresholds<br>- classify the attributes as quasi-identifiers, sensitive attributes, and non-sensitive attributes<br>- determine the actual dimension of QIDs that should be used in an anonymisation operation that will achieve optimum case | - data preprocessing<br>- compute risk rate<br>- select classification thresholds<br>- classify the dataset attributes into quasi-identifiers, sensitive attributes, and non-sensitive attributes<br>- echo chamber optimisation |
| Performance measures | - privacy gain<br>- non-uniform entropy | - average equivalent class size metric<br>- discernibility metric<br>- normalised certainty penalty |
| Key takeaways | - accurate identification of QIDs is an important issue for the success and validity methods of privacy-preserving outsourced data that seek to avoid privacy leakage caused by QID linking<br>- the proposed identification algorithm has better performance and is more perfect in terms of privacy provided against data utility when compared with other works | - the developed optimized clustering-based algorithm with the privacy preservation model extensively minimizes the leakage of private information and the utilisation of data is well-maintained compared with other existing algorithms |

| | | | | | |
|---|---|---|---|---|---|
| **What are the methods used to evaluate data usefulness of an anonymised dataset?** | | | | | |
| Title | Privacy protection in social science research: possibilities and impossibilities [29] | An experimental comparison of quality models for health data de-identification [30] | Utility-driven assessment of anonymized data via clustering [31] | Utility-preserving transaction data anonymization with low information loss [32] | A generic method for assessing the quality of de-identified health data [33] |
| Authors | Albright JJ | Eicher J, Kuhn KA, Prasser F | Ferrão ME, Prata P, Fazendeiro P | Loukides G, Gkoulalas-Divanis A | Prasser F, Bild R, Kuhn KA |
| Publication year | 2011 | 2017 | 2022 | 2012 | 2016 |
| Objectives | - contribute to an understanding of the technical issues involved with SDC | Answer the following questions:<br>- How do common models for measuring data quality influence the way in which datasets are transformed?<br>- If different models are used, how are the obtained results related to each other?<br>- How well is de-identified data, obtained by using different quality models, suited for real-world applications? | - proposal to adjust the utility model to the research question in the applied field of study as complementary to data utility quantified by standard metrics, no matter the substantive applied field of study<br>- provide insight into the differences between anonymised and original datasets and debate its relevance for research purposes | - propose a novel approach for anonymising data in a way that satisfies data publishers' utility requirements and incurs low information loss | - development of a generic variant to non-uniform entropy which can be used to assess the information loss induced by transforming data with arbitrary combinations of full-domain generalisation, local recoding and record or value suppression |
| Methodology | - introduction of the field of SDC by defining key terms, describing how researchers quantify risk, identifying options to minimise risk, and outlining how these decisions affect the usefulness of a data file<br>- description of the implications of SDC for political science research, namely the problems it introduces for variance estimation in complex surveys<br>- outline where the field of SDC is headed | The used quality models:<br>- Average Equivalence Class Size (AECS)<br>- discernibility<br>- precision<br>- loss<br>- ambiguity<br>- Kullback-Leibler (K.-L.) divergence<br>- non-uniform entropy | - clustering as an utility indicator | - introduction of Utility Criterion (UC), a measure that can quantify data utility under different generalisation models and be employed by effective anonymisation algorithms<br>- development of a novel anonymisation algorithm<br>- experimental evaluation of the approach using two datasets | - non-uniform entropy<br>- a generic variant to non-uniform entropy |
| Key takeaways | - disclosure risk may be higher than researchers realise<br>- the proactive steps data collection organisations take to minimise disclosure risk can affect the ability of the end user to accurately estimate statistical relationships | - different models are suited best for different application scenarios<br>- the non-uniform entropy model provides the best results for general purpose usage | - when working with low dimensionality datasets, no matter the method of anonymisation, the results obtained suggest that the replacement of original data by their anonymised versions may jeopardise the proper data analysis, the data-based inferences or deductions and even the conclusions of the scientific research | - the UAR anonymisation algorithm incurs significantly lower information loss than the state-of-the-art methods | - the used method provides a unified framework in which this model can be used to assess and compare the quality of differently transformed data to find a good or even optimal solution to a given de-identification problem |

## Appendix 6: Filled out quality assessment checklists

**Given the already present definitions entailing quasi-identifiers and sensitive attributes, how can the identification process for these types of attributes be standardised using an algorithmic approach?**

| Study | Quasi-identifier recognition algorithm for privacy preservation of cloud data based on risk reidentification [22] | Quasi-identifier recognition with echo chamber optimization-based anonymization for privacy preservation of cloud storage [21] |
|---|---|---|
| Was there a clear description of the aims and purposes of the research? | Yes | Yes |
| Was the algorithm clearly described (e.g. flowchart, pseudocode, ...) | Yes | Yes |
| Was the experimental dataset described? | Yes | Yes |
| Were any metrics used to validate the methodology? | Yes | Yes |
| Was the quality of the anonymised data assessed? | Yes | Yes |
| Was the quality assessment done using simple statistical methods or machine learning? | Statistical methods □ no | Statistical methods □ no |
| Were there any hyperparameters that were finetuned? | No | No |
| Is there a repository of the code? | No | No |
| **Final score** | **5/8** | **5/8** |

**What are the methods used to evaluate data usefulness of an anonymised dataset?**

| Study | Privacy protection in social science research: possibilities and impossibilities [29] | An experimental comparison of quality models for health data de-identification [30] | Utility-driven assessment of anonymized data via clustering [31] | Utility-preserving transaction data anonymization with low information loss [32] | A generic method for assessing the quality of de-identified health data [33] |
|---|---|---|---|---|---|
| Was there a clear description of the aims and purposes of the research? | No | Yes | Yes | Yes | Yes |
| Was the experimental dataset described? | No | Yes | Yes | Yes | Yes |
| Were any metrics used to validate the methodology? | No | No | No | No | No |
| Were the metrics clearly described? | No | Yes | Yes | Yes | Yes |
| Is there a repository of the code? | No | No | Yes | No | No |
| **Final score** | **0/5** | **3/5** | **4/5** | **3/5** | **3/5** |

# Appendix 7: Classification of attributes

| Classification of attributes in the 500 row dataset | | |
|---|---|---|
| **Column name** | **Re-identification risk (%)** | **Classification (α = 25.0, β = 1.0)** |
| **bmi** | 38.50 | Sensitive attribute |
| **ms_diagnosis_date** | 27.65 | Sensitive attribute |
| **edss** | 22.58 | Quasi-identifier |
| **age** | 5.49 | Quasi-identifier |
| **comorbidities** | 3.63 | Quasi-identifier |
| **covid19_symptoms** | 3.12 | Quasi-identifier |
| **ms_type** | 1.34 | Quasi-identifier |
| **covid19_ventilation** | 0.96 | Non-sensitive attribute |
| **covid19_outcome_recovered** | 0.61 | Non-sensitive attribute |
| **covid19_icu_stay** | 0.53 | Non-sensitive attribute |
| **covid19_confirmed_case** | 0.50 | Non-sensitive attribute |
| **report_source** | 0.40 | Non-sensitive attribute |
| **sex** | 0.40 | Non-sensitive attribute |
| **covid19_admission_hospital** | 0.40 | Non-sensitive attribute |
| **covid19_diagnosis** | 0.40 | Non-sensitive attribute |

| Classification of attributes in the 1000 row dataset | | |
|---|---|---|
| **Column name** | **Re-identification risk (%)** | **Classification (α = 10.0, β = 1.0)** |
| **bmi** | 20.98 | Sensitive attribute |
| **ms_diagnosis_date** | 13.81 | Sensitive attribute |
| **edss** | 10.04 | Sensitive attribute |
| **age** | 2.66 | Quasi-identifier |
| **comorbidities** | 1.80 | Quasi-identifier |
| **covid19_symptoms** | 1.54 | Quasi-identifier |
| **ms_type** | 0.67 | Non-sensitive attribute |
| **covid19_ventilation** | 0.52 | Non-sensitive attribute |
| **covid19_outcome_recovered** | 0.31 | Non-sensitive attribute |
| **covid19_icu_stay** | 0.26 | Non-sensitive attribute |
| **covid19_confirmed_case** | 0.26 | Non-sensitive attribute |
| **report_source** | 0.20 | Non-sensitive attribute |
| **sex** | 0.20 | Non-sensitive attribute |
| **covid19_admission_hospital** | 0.20 | Non-sensitive attribute |
| **covid19_diagnosis** | 0.20 | Non-sensitive attribute |